# Table of Contents Graphic

# Crossover from Kondo assisted suppression to co-tunneling enhancement of tunneling magnetoresistance via ferromagnetic nanodots in MgO tunnel barriers

Hyunsoo Yang, See-Hun Yang and Stuart Parkin

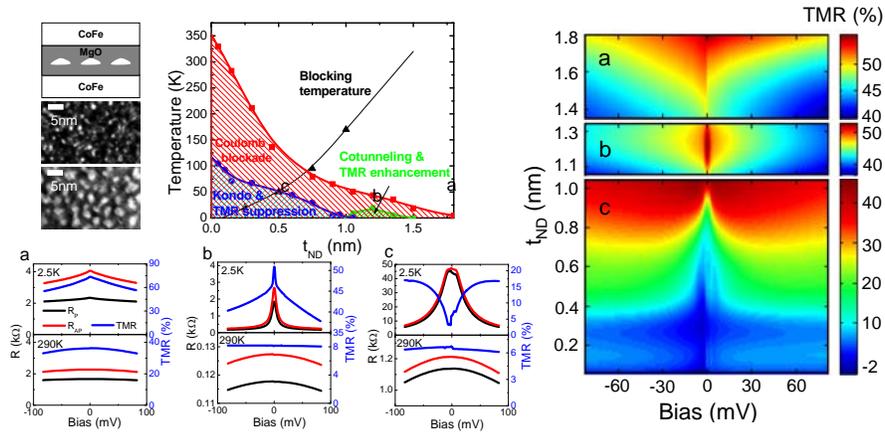

# Crossover from Kondo assisted suppression to co-tunneling enhancement of tunneling magnetoresistance via ferromagnetic nanodots in MgO tunnel barriers

*Hyunsoo Yang, See-Hun Yang and Stuart Parkin*[*]

IBM Almaden Research Center, 650 Harry Road, San Jose, California 95120, USA

## Abstract

**The dependence of the tunneling magnetoresistance (TMR) of planar magnetic tunnel junctions on the size of magnetic nanodots incorporated within MgO tunnel barriers is explored. At low temperatures, in the Coulomb blockade (CB) regime, for smaller nanodots the conductance of the junction is *increased* at low bias consistent with Kondo assisted tunneling and the TMR is *suppressed*. For slightly larger nanodots but within the CB regime, the TMR is *enhanced* at low bias, consistent with co-tunneling.**

Magnetic tunnel junctions (MTJ) exhibit giant magnetoresistance in small magnetic fields which arises from the flow of spin polarized current through an ultra thin tunnel barrier separating two magnetic electrodes. The current through an MTJ device depends on the magnetic orientation of the electrodes and is typically higher when the electrode moments are parallel (P) than when they are antiparallel (AP)[1]. It has recently been demonstrated that the spin polarization of the tunneling current can be greatly enhanced by using crystalline tunnel barriers formed from MgO as compared with conventional amorphous barriers formed from alumina, due to spin filtering across the MgO layer[2-6]. The magneto-transport properties of magnetic granular alloys and magnetic tunnel junction devices with magnetic nanodots embedded in amorphous dielectric matrices[7, 8] and tunnel

---

[*] To whom correspondence should be addressed: email: parkin@almaden.ibm.com; phone: (408) 927 -2390

barriers[9, 10], respectively, have been studied by several groups but no systematic studies of the dependence on these properties on the nanodot size have been made.

Here we show that a wide range of magneto-transport properties in magnetic tunnel junctions with barriers containing magnetic nanodots can be accounted for by Kondo-like physics and that as the size of these nanodots is systematically increased the tunnelling properties change from those characteristic of Kondo tunnelling which suppresses the TMR near zero bias, to those of correlated tunneling which increases the TMR at low bias.

The MTJ device (see Figure 1a) is composed of CoFe ferromagnetic (F) electrodes, and an MgO tunnel barrier [2]. Both F electrodes are exchange biased but the exchange bias field is designed to be significantly stronger for the lower electrode so that the moment of each electrode can be switched independently. A thin CoFe layer of nominal thickness $t_{ND}$ is inserted in the middle of the MgO layer. When this layer is thinner than ~2 nm it forms a discontinuous layer of nanodots, as revealed by plan-view transmission electron microscopy (TEM). This means that the diameter of the nanodots is considerably larger than the nominal layer thickness. For example, when $t_{ND}$= 0.25 and 0.75 nm, the nanodot diameters are estimated to be ~1.53±0.4 nm and ~3.2±0.7 nm, respectively (see Figure 1b and 1c, and the Supplementary Information). The chemical state of individual nanodots was examined by electron energy loss spectroscopy in a high resolution scanning TEM which revealed clear evidence for the presence of both metallic Co and Fe.

The magneto-transport properties of the MTJs are considerably affected by the presence of the nanodots in the MgO tunnel barrier, as shown in Figure 2. In particular, as the temperature is reduced, the resistance of the MTJ increases, but at a greater rate, the smaller the nanodot size. This is due to a Coulomb blockade (CB) effect[11], which has previously been seen in MTJs formed with discontinuous Co and Fe layers within alumina[10, 12-14] and MgO[15] tunnel barriers. At high temperatures and bias voltages the tunneling is dominated by sequential tunneling through the nanodots since the barrier is very thick (thus direct tunneling across the MgO barrier is small). The nanodots are so small that there is a significant increase in energy when an electron tunnels onto a dot which, at low temperature and bias compared to this energy, thereby depresses the tunneling conductance across the MTJ device[9].   By taking into account both indirect and direct tunneling mechanisms (following ref. [16], as discussed in detail in the Supplementary Information) , the temperature dependence of the tunneling resistance can be well accounted for, as shown by the solid curves in Figure 3a.   From these fits a characteristic temperature $T_{CB}$, below which CB effects are important, can be derived.   $T_{CB}$ is proportional to the charging energy needed to add a single electron to the dot (see the Supplementary Information).  CB effects are observed when $t_{ND}$ <1.8nm with $T_{CB}$ increasing to ~330 K when $t_{ND}$ =0.05 nm (see Figure 1d).   The calculated CB energy is consistent with the average size of the nanodots inferred from plan-view TEM micrographs.  For example, when $t_{ND}$ = 0.45nm, the CB energy is ~53 meV which gives a nanodot diameter of ~2.6 nm (see the Supplementary Information).

The nanodots are so small that they are superparamagnetic[17] which means that the direction of the magnetic moment of an individual nanodot fluctuates due to thermal fluctuations above a characteristic blocking temperature $T_B$.  By contrast with the CB effect, $T_B$ decreases as the nanodot volume is decreased.   $T_B$ was measured from the magnetic properties of multilayers of the form [3nm MgO/$t_{ND}$ CoFe]$_{20}$ since the magnetic signal from a single nanodot layer is too weak to be measured alone by SQUID magnetometry (Figure S2).   $T_B$ was determined from the temperature at which the zero field cooled (ZFC) magnetization shows a maximum value [18].   The dependence of $T_B$ on $t_{ND}$ is shown in Figure 1d.  $T_B$ is reduced below the CB temperature when $t_{ND}$ is smaller than ~0.7nm.    The ZFC magnetization data show clear evidence for blocking effects with the magnetization dropping below $T_B$ to nearly zero at 5 K.  However, field cooled (FC) data show evidence for a considerable dispersion in the size of the nanodots so that many nanodots remain unblocked with fluctuating magnetic moments even for temperatures well below $T_B$.  Interestingly, as $t_{ND}$ is increased the FC data show that an increasing proportion of the nanodot moments are blocked so that above $t_{ND}$~1nm most nanodot moments are blocked below $T_B$ (see Figure S2 in the Supplementary Information).

A second consequence of the CB phenomenon arising from the nanodots is a strongly enhanced bias voltage dependence of the tunneling resistance in the CB regime (i.e. when $T<T_{CB}$) when the bias voltage is below the CB charging energy[14]. The bias voltage dependence of the resistance of an MTJ with no nanodot layer and nanodot layers 0.45nm and 1.2nm thick are compared in Figure 2 for temperatures ranging from 2.5 to 300 K. The resistance of the pristine MTJ, with its moments either parallel $R_P$ or antiparallel $R_{AP}$, shows a weak dependence on the bias voltage V (Figure 2a) for all temperatures. By contrast, at low temperatures, $R_P$ and $R_{AP}$ decrease rapidly with V when the nanodot layer is present, the more rapidly the bigger the nanodot. For the examples shown in Figure 2, at 2.5 K the resistance is decreased by half at V~ 26 mV and 7 mV, when $t_{ND}$ = 0.45 and 1.2nm, respectively, for both the P and AP states. These voltages scale with $T_{CB}$ (see Figure 1d) and are clearly related to the CB effect.

An important result is a strong suppression of the TMR at low temperatures and bias voltages in the CB regime when the nanodot layer is thinner than ~1 nm (Figure 2e and 2f). A similar phenomenon is observed for nanodot layers even as thin as 0.05 nm (see Figure 4). However, there is a small region of nanodot layer thickness, when $t_{ND}$ is slightly thicker than ~1nm, but still within the CB regime, where we observe an enhancement of the TMR at low bias voltage and temperature (see Figure 1d and 2i). This latter phenomenon has recently been observed in large-area MTJs with alumina tunnel barriers in which a nanodot layer formed from 1nm thick $Co_{90}Fe_{10}$ was inserted[10] and was attributed to co-tunneling[9]. Co-tunneling is the correlated tunneling of electrons when sequential uncorrelated tunneling through the magnetic nanodots is suppressed for voltages below the CB charging energy $E_{CB}$[9]. Our observation of suppressed TMR for smaller nanodots was not predicted by this theory which, however, did not consider spin-flip processes on the nanodots.

The suppression of TMR is consistent with theoretical models in which the spin-dependent tunneling through nominally non-magnetic nanodots in the CB regime between ferromagnetic electrodes is considered[19, 20]. In these models the nanodots are so small that they contain a finite number of electrons and thus, in the CB regime when there are an odd number of electrons on the nanodot, the dot has a finite spin. The screening of the nanodot spin through a Kondo interaction with the spin-polarized electrons in the ferromagnetic electrodes was predicted to lead to a suppression of the TMR when $eV<E_{CB}$[21]. This effect has been observed in a very small number of samples in experiments in which tunneling via individual (or possibly several) non-magnetic $C_{60}$ molecules between Ni electrodes was studied in break junctions formed by electromigration of a Ni point contact[22]. Here, we appear to observe a similar effect in planar macroscopic tunnel junctions in which tunneling takes place through many magnetic nanodots of a considerable size. For example, when $t_{ND}$=0.45nm, we estimate that the nanodot contains ~600 atoms. Moreover, the CoFe nanodots have large magnetic moments - each atom has a moment at 2.5 K of ~1.8 $\mu_B$. Kondo effects are usually considered to be important only when the magnetic moment being screened is small[23], although recent theoretical work has suggested that Kondo effects may be possible with large magnetic moments when the magnetic moments have considerable magnetic anisotropy[24, 25]. In our case the magnetic nanodots are pancake-shaped (see the Supplementary Information) which will lead to significant magnetic anisotropy.

There are several other characteristic features of Kondo assisted tunneling, which we observe in MTJs with nanodots. One of these is a peak in the tunneling conductance versus bias voltage curve centred at zero voltage which has a characteristic width proportional to the strength of the Kondo interaction i.e. the Kondo temperature $T_K$. The peak can be accounted for by a Kondo resonant tunneling channel. Conductance versus voltage curves are compared in Figure 3b for MTJs with $t_{ND}$=0 and 0.5 nm. The MTJ without nanodots shows a dip in conductance near zero bias whereas MTJs with nanodot layers, for thicknesses ranging from 0.05 – 1nm, show a conductance peak. From the half-width at half maximum $\Delta V$ of the conductance peak we can estimate $T_K$ ~55 K from the relationship $\Delta V=k_B T_K/e$[26, 27] with $t_{ND}$=0.5 nm. We estimated $T_K$ using two other methods. Firstly, we considered the temperature dependence of the zero bias conductance peak. Since there is a considerable background to the conductance versus voltage curve due predominantly to the increasing number of available states in the receiving F electrode at higher bias voltages, we fitted the conductance curves at each temperature for voltages from ~2 $\Delta V$ to 50 mV with a second order polynomial and subtracted this curve from the measured curve to calculate the enhanced conductance at zero bias $\Delta G_K/G_B$ relative to the fitted background conductance at zero bias $G_B$[28]. The temperature dependence of $\Delta G_K/G_B$, which is shown in Figure 3c for $t_{ND}$=0.2 and 0.5 nm, can be well described by an empirical formula developed to account for the temperature dependence of the Kondo conductance in a single electron transistor[29], and which has been widely used to describe the behaviour in a variety of Kondo systems[30, 31]. For the case of $t_{ND}$=0.2 nm and for the P configuration, $T_K$, which we extract from this fit, is ~76K, which compares well with the Kondo temperature derived from the conductance peak width at 2.5 K of ~ 69 K. Finally, the Kondo temperature can also be estimated, although more approximately, from the temperature at which the TMR is no longer suppressed at low bias voltage. For example, from the detailed dependence of TMR on voltage and temperature for the case of $t_{ND}$~0.45nm shown in Figure 2e, we estimate $T_K$~60 K. We note that values of $T_K$ are systematically higher by about 10-20% for the AP configuration.

The dependence of the Kondo temperature, derived from averaging the values found from the three methods discussed above, is plotted as a function of the nanodot layer thickness in Figure 1d. For the thinnest nanodot layer $T_K$ ~ 100 K, a very high Kondo temperature, but $T_K$ decreases with increasing thickness to zero when $t_{ND}$ is ~ 1nm thick. From Figure 1d it is clear that $T_K$ is always significantly lower than $T_{CB}$, consistent with theoretical expectations that Kondo assisted tunneling will only be found within the CB regime[21]. The high Kondo temperature might be indicative of a strong interaction between the ferromagnetic electrodes and the nanodots through the highly oriented MgO(100) tunnel barriers. Indeed, evidence for significant antiferromagnetic coupling of magnetic electrodes through MgO(100) barriers has been found[32, 33].

We note that the peak conductance near zero bias in the Kondo regime is several orders of magnitude below that calculated within the unitary limit. There are many possible reasons to account for this observation including, the lack of tunability of the energy of the quantized electronic levels on the nanodots[34], which will be very sensitive to the nanodot size and shape[35] and their magnetic configuration, interference within multiple conductance channels through individual nanodots, and asymmetry in the tunnelling barriers[36].

The detailed dependence of the voltage dependence of the tunneling magnetoresistance is plotted in Figure 4 as a function of the nanodot layer thickness at 2.5K. We attribute the overall suppression of the maximum TMR observed as $t_{ND}$ is decreased to the reduced blocking temperature and the larger magnetic anisotropy of the smaller nanodots. This means that increasingly large magnetic fields are required to fully align the magnetic moments of the nanodots parallel to each other when the MTJ device is cooled. Since each device was cooled in a fixed field of 1T the degree of magnetic alignment is progressively reduced as $t_{ND}$ is decreased and the nanodots become smaller. The data in Figure 4c show that as $t_{ND}$ is increased from 0.05 to ~1nm the relative reduction of TMR at zero bias from its maximum value at each $t_{ND}$ becomes increasingly smaller (see Figure S4). The suppression of TMR is such that in many cases, and always for the very smallest nanodots, the TMR becomes negative i.e. the conductance becomes greater for the AP alignment of the ferromagnetic electrodes than for the P alignment (see Figure 3b). This inversion of the sign of the TMR is attributed to an exchange of the electron's spin as it tunnels via a virtual state on the nanodot. This is consistent with a Kondo phenomenon[23, 34]. The enhancement of TMR at low bias seen for nanodot layers with thicknesses in the narrow range of ~1-1.4 nm can also be clearly seen in Figure 4b (see also Figure 2h). Outside the CB regime the TMR has a weak dependence on voltage (below ±100 mV, the range shown in Figure 4a).

It is interesting to note that the crossover from a Kondo like tunneling phenomenon with enhanced conductance and suppressed TMR to a regime of depressed conductance and enhanced TMR at low bias occurs at a nanodot layer thickness of ~1 nm. This is the same thickness below which field cooled magnetization data indicates a dispersion of nanodot blocking temperatures and considerable fluctuations of nanodot magnetic moments persisting to low temperatures, and above which there is a single nanodot blocking temperature. Thus the crossover from Kondo to co-tunneling behaviour is likely correlated with suppression of the fluctuations of the nanodot magnetic moments. Although the tunnelling characteristics for the smallest nanodots can be described by Kondo physics, the possibility of observing Kondo phenomenon involving quantum dots with large magnetic moments requires a stronger theoretical foundation and remains an open question.

**Acknowledgements.** We thank Teya Topuria and Phil Rice for TEM data and analysis. We thank DMEA for partial support of this work. We gratefully acknowledge useful discussions with Profs. Sadamichi Maekawa, Jan Martinek and David Goldhaber-Gordon.

**Supplementary Information Available:** Sample preparation and measurement; Definition of the Coulomb blockade temperature; Estimate of the nanodot size and shape; Superparamagnetic behaviour of nanodots; Influence of nanodot magnetic orientation on the TMR suppression. This material is available free of charge via the Internet at http://pubs.acs.org.

**Figure 1.** (a) Illustration of a magnetic tunnel junction (MTJ) device with a discontinuous layer deposited in the middle of the tunnel barrier, forming a layer of nanodots. (b)(c) Plan view scanning transmission electron micrographs for $t_{ND}$=0.25 nm and $t_{ND}$=0.75 nm in related structures in which the top electrode is not grown but is substituted by 2 nm Mg and the bottom electrode is milled away to leave only the layer of CoFe nanodots sandwiched between MgO. White regions correspond to nanodots while black regions indicate MgO. (d) Phase diagram showing the dependence of the blocking temperature $T_B$, the Kondo temperature $T_K$ and the Coulomb Blockade (CB) temperature on $t_{ND}$. The CB region is shaded in red. Within the blue region the TMR is suppressed and the conductivity enhanced near zero bias, consistent with Kondo assisted tunneling. By contrast within the smaller green region the TMR is enhanced and the conductivity reduced near zero bias, consistent with co-tunneling. Solid lines through the data points are guides to the eye. Note that $T_B$ corresponds to the blocking temperature of the largest nanodots (see the Supplementary Information) so that smaller nanodots will have lower blocking temperatures.

**Figure 2.** The temperature and bias voltage dependence of the dc resistance and tunneling magnetoresistance (TMR) of a MTJ without the nanodot layer (a-c) and with a nanodot layer in (d-i). Data are shown for the resistance when the magnetic moments of the upper and lower exchange-biased ferromagnetic electrodes are parallel (P) and antiparallel (AP). The nanodot layer thickness is 0.45 nm in (d-f) and 1.2 nm in (g-i).

**Figure 3.** (a) The temperature dependence of the dc MTJ resistance at zero bias interpolated from current versus bias voltage curves. Examples are shown for four nanodot layer thicknesses indicated in the Figure. Filled (open) symbols correspond to $R_P$ ($R_{AP}$) and solid lines are fits to the $R_P$ data, as discussed in the Supplementary Information. (b) Differential conductance versus bias voltage curves at 2.5 K for the P and AP magnetic configurations for an MTJ without a nanodot layer (solid and open black squares) and an MTJ with a nanodot layer of thickness 0.5 nm (solid and open red circles). $G_K$ (open black dot) is the measured conductance at zero bias, and $G_B$ (solid black dot) is the conductance at zero bias interpolated by fitting the curves at higher voltages. (c) Temperature dependence of the conductance enhancement ratio $(G_K-G_B)/G_B$ for MTJs with $t_{ND}$= 0.2 and 0.5 nm in the P and AP configurations. The solid lines through the data are fits using the formula[29] $\Delta G_K(T)/G_B = G_0/1(1+(2^{1/s}-1)T^2/T_K^2)^s$ where $T_K$ =31 K (s=1.58) and $T_K$ =36 K (s=0.98) for the P and AP states, respectively, for the case of $t_{ND}$=0.5 nm. For $t_{ND}$ = 0.2nm, $T_K$ =76 K (s=2.19) and $T_K$ =81 K (s=1) for the P and AP states, respectively.

**Figure 4.** The bias voltage dependence of dc TMR at 2.5 K as a function of the nanodot layer thickness. The contour plots are created by interpolation between the experimental data points. In order to enhance the visibility of the TMR peaks (b) and dips (c) the plot is divided into 3 parts: (a) sequential tunnelling ($t_{ND}$=1.35-1.8 nm), (b) co-tunneling ($t_{ND}$=1-1.35 nm), and (c) Kondo-assisted tunneling ($t_{ND}$=0.05-1 nm) regimes. The color represents different TMR scales in each part.

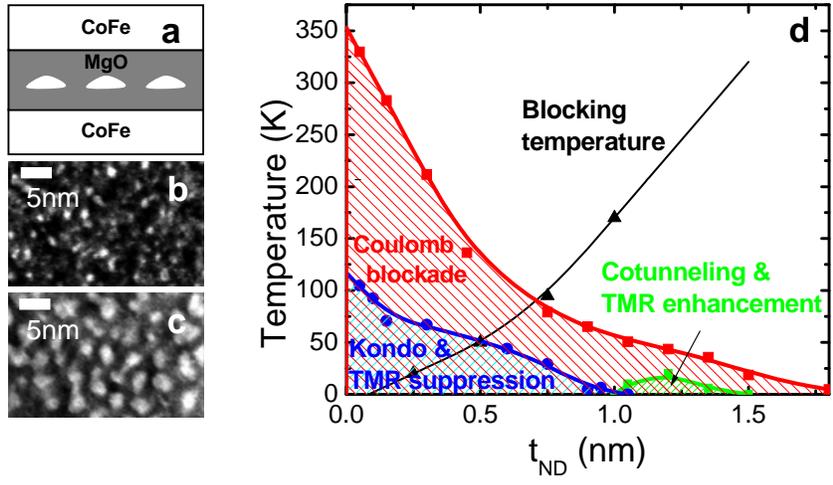

Figure 1

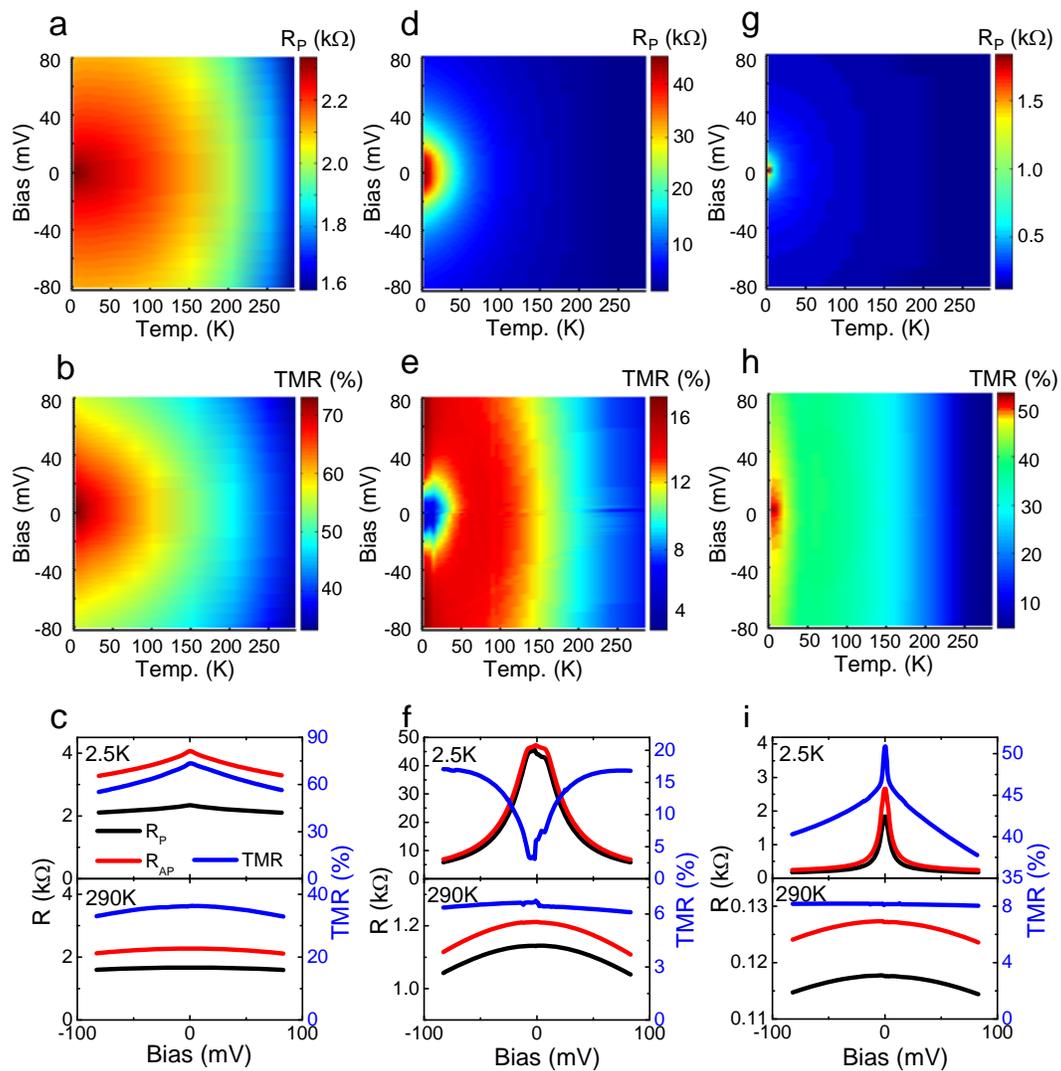

Figure 2

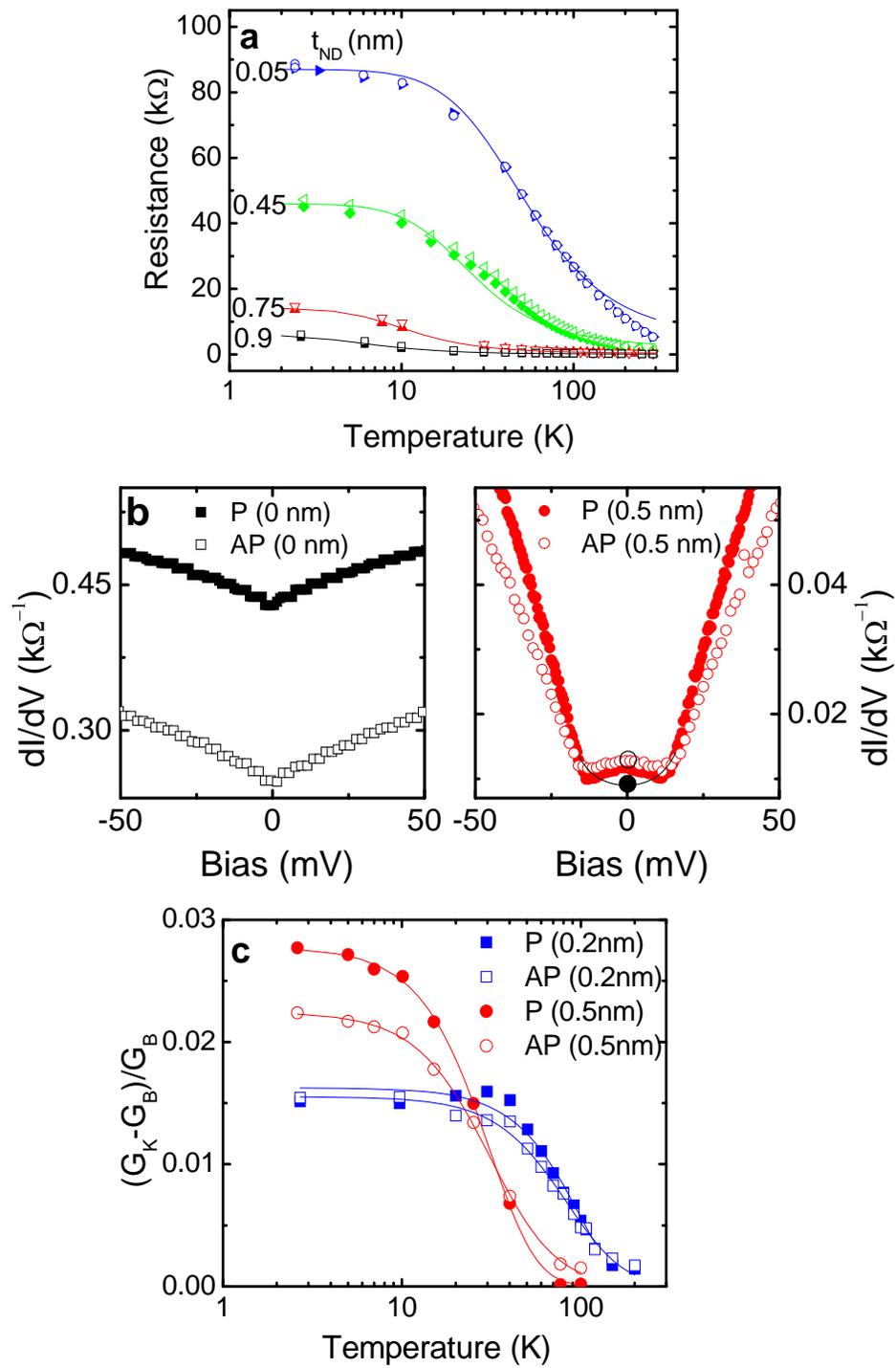

Figure 3

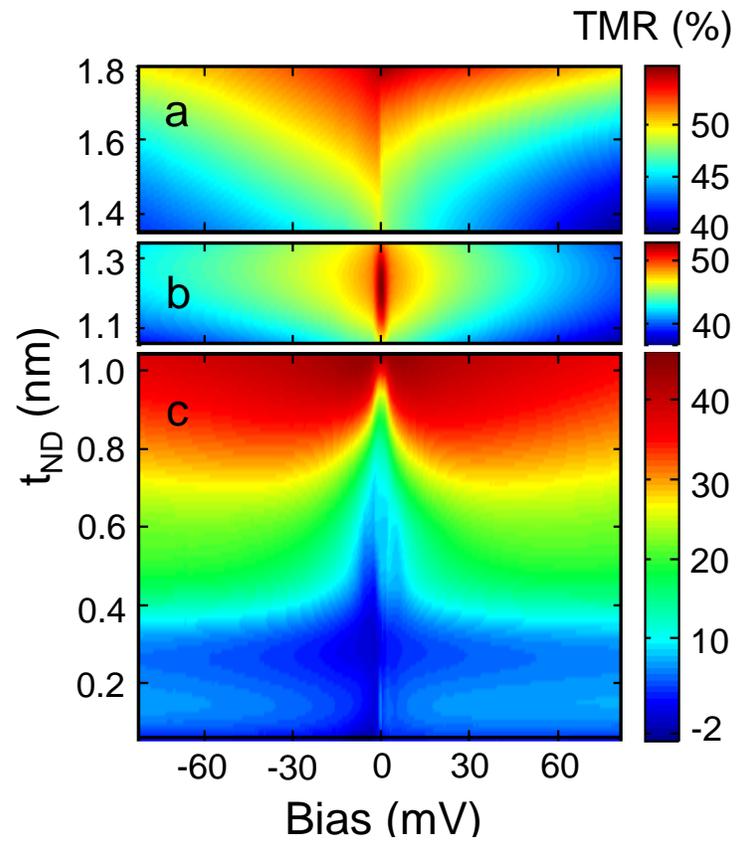

Figure 4

# Crossover from Kondo assisted suppression to co-tunneling enhancement of tunneling magnetoresistance via ferromagnetic nanodots in MgO tunnel barriers

**Supplementary Information**

Hyunsoo Yang, See-Hun Yang and Stuart Parkin

*IBM Almaden Research Center, 650 Harry Road, San Jose, California 95120, USA*

**Sample preparation and measurement**

MTJs were deposited on 25 nm $SiO_2$/Si through a sequence of metal shadow masks using dc magnetron sputtering at room temperature. MTJs without a nanodot layer were formed from 10 Ta / 25 $Ir_{22}Mn_{78}$/ 3.5 $Co_{70}Fe_{30}$/ 0.8 Mg/ 2.8 MgO/ 7 $Co_{70}Fe_{30}$/ 10 Ta (layer thicknesses in nm). MTJs with a nanodot layer were comprised of 10 Ta/ 25 $Ir_{22}Mn_{78}$/ 3.5 $Co_{70}Fe_{30}$/ 0.8 Mg/ 2.5 MgO/ $t_{ND}$ $Co_{70}Fe_{30}$/ 0.8 Mg/ 2.5 MgO/ 7 $Co_{70}Fe_{30}$/ 15 $Ir_{22}Mn_{78}$/ 10 Ta, (thicknesses in nm). The junction area is ~700μm×700μm. Magnetic fields of +10,000 and -500 Oe were applied to set the state of the MTJ in the P and AP states, respectively. The TMR is defined as $(R_{AP}-R_P)/R_P$, where $R_{AP}$ and $R_P$ are the dc resistances in the AP and P configurations, respectively. The resistance of the shadow-masked MTJs was measured using a standard four-probe method.

**Definition of the Coulomb blockade temperature inferred from fits of the temperature dependence of the tunneling resistance**

Following Giaever and Zeller we assume that the electrons tunnel from the electrodes onto the magnetic nanodots with a probability given by $\exp(-E_C/k_BT)$ where $E_C$ is the activation energy of the dot. We assume a log-normal distribution of nanodot diameter with fitting parameters given by the mean nanodot diameter and distribution width. We also assume that the tunneling probability is proportional to the area of the dot and we find the tunneling current at a given temperature by integrating over the dot distribution. In addition to the tunneling conductance via the dots we also include a temperature independent conductance which represents direct tunneling through the insulating matrix. This term dominates at low temperatures when the conductance through the nanodots becomes very small due to the CB effect. We find that very reasonable fits to the temperature dependence of our device resistance are obtained with reasonable values of fitted dot diameter and distribution width. The activation energy is also a fitting parameter from which we derive the Coulomb Blockade temperature as discussed in the Supplementary Information.

The temperature dependence of the tunneling resistance R was fitted with an equation (1) of the form $1/R = \alpha/R_1 \times \int_0^\infty n(E_C) f_D(E_C) e^{-E_C/kT} dE_C + 1/R_2$ where $\alpha$ is the normalization coefficient $1/\int_0^\infty n(E_c) f_D(E_c) dE_c$, $E_C$ the activation energy, $n(E_C)$ the area of nanodots, $f_D(E_C)$ the nanodot size log-normal distribution function given in the next section (ref. Zeller and Giever). The first term describes indirect thermally activated tunnelling via the metallic nanodots in the tunnel barrier and the second $1/R_2$ term describes direct tunnelling across the barrier.

The charging energy is given by the equation (2) $E_{CB} = (2\pi^2 h^2/m\phi)^{0.5} E_C/s$, where m is the tunneling electron mass, $\phi$ is the effective tunnel barrier height, and s is the average spacing between the nanodots and the electrodes, and $E_{CB}$ is the characteristic Coulomb blockade (CB) energy of a single nanodot (ref. 15). Note that the activation energy is proportional to the Coulomb charging energy $E_{CB}$. The activation energy and $R_1$ and $R_2$ are extracted from the fit. The Coulomb blockade temperature $T_{CB}$ is then found from the relation $T_{CB} = E_C/k_B$.

For the case of $t_{ND} = 0.45$ nm, a fit to the resistance versus temperature curve shown in Figure 3a using equation (1) gives $E_C = 11.7$ meV, $R_1 = 2.05$ k$\Omega$ and $R_2 = 46$ k$\Omega$. The data is well described by equation (1). This activation energy corresponds to $T_{CB} = 136$ K. The dependence of $T_{CB}$ on $t_{ND}$ is shown in Figure 1D.

Using reasonable values of the tunnel barrier height ($\phi \sim 3$ eV) and the tunneling electron effective mass ($m \sim 0.2 m_0$, where $m_0$ is the free electron mass), the CB energy ($E_{CB}$) is found to be

~53meV from equation (2). Note that the activation energy is considerably smaller than the CB energy. This has also been reported by Tran *et al.* (S1) for the case of tunneling through gold nanoparticle multilayers inside the CB regime.

**Estimate of the nanodot size and shape**

The average diameter of the nanodots (*d*) can be estimated from the magnitude of the CB energy which is given by $E_{CB}=e^2/2C$ where C is the capacitance of a single nanodot. *C* is given by $\pi\epsilon\epsilon_0 d(d+4s)/2s$ for the case of a spherical nanodot of diameter *d* positioned a distance *s* from each of the two metallic F electrodes (ref. 11), where $\epsilon$ is the dielectric constant ~8 in sputtered MgO thin films (S2), $\epsilon_0=8.85\times 10^{-12}$ F/m, and $s \sim 2.2$ nm. The diameter of the nanodots (*d*) is then estimated to be 2.6 nm for the case of $t_{ND} = 0.45$ nm.

The size of the nanodots can also be estimated from their magnetic blocking temperature (see below). As discussed in the main text the blocking temperature was measured for different sized nanodots from the superparamagnetic properties of related nanodot multilayers. For the case of $t_{ND}$ = 0.5 nm, the measured blocking temperature of ~40K is consistent with $d\sim 2.5$ nm (ref. 19) assuming that the moments of all the atoms in individual nanodots are largely ferromagnetically aligned with each other and that each nanodot behaves like a magnetic particle with a well defined macroscopic magnetic moment. Thus these studies show that the magnetic moments of the nanodots are not rigid but will fluctuate due to thermal fluctuations or induced by the tunneling of spin polarized electrons.

Finally, the diameter and height of the nanodots was estimated from through foil TEM images (Figure 1b and 1c). The distribution of the diameters of the nanodot array, calculated from the TEM image for the case of $t_{ND} = 0.25$ nm (Figure 1b), is shown in Figure S1. The distribution was found by measuring the diameter of 100 nanodots and averaging 2 measurements per nanodot. The distribution is well described by a log-normal function of the form $f_D(d) = N/(\sigma d\sqrt{2\pi})\exp[-\ln^2(d/d_m)/(2\sigma^2)]$ whose mean is $d_m\sim 1.48$ nm and standard deviation is 0.2. The average diameter of the nanodots, given by $d_{avg} = \frac{1}{n}\sum_{i=1}^{n} d_i f_D(d_i)$, is ~1.53 ± 0.4 nm. For the case of $t_{ND}$ = 0.75 nm, the average diameter of the nanodots was estimated to be ~3.2 ±0.7 nm.

The height of the nanodots was inferred by estimating the fraction of the MgO layer covered by the nanodots. For $t_{ND}$ = 0.25nm and 0.75nm, respectively, the coverage of the MgO layer is estimated to be 25% and 50%. Therefore, assuming the dots have a uniform height, their heights are estimated to be roughly four and two times thicker than the deposited CoFe layer thickness, $t_{ND}$, giving heights of ~1nm and 1.5nm, respectively. Thus the aspect ratio of the nanodots has a weak dependence on $t_{ND}$ with a height to diameter aspect ratio of ~0.65 (1/1.53) for $t_{ND}$= 0.25 nm and ~0.47 (1.5/3.2) for $t_{ND}$=0.75 nm, as estimated from the TEM studies.

**Superparamagnetic behaviour of nanodots**

Zero field cooled (ZFC) and field cooled (FC) magnetization versus temperature curves are shown in Figure S2 for nanodot multilayers of the form [MgO / $t_{ND}$ CoFe]$_{20}$ for $t_{ND}$ = 0.25, 0.5, 0.75 and 1nm. The ZFC and FC data were taken by first cooling the sample to 5K in zero field and 1T, respectively. The sample magnetization was then measured in 0.05 T as the sample temperature was increased from 5K to 300K. As the temperature is increased the ZFC magnetization increases and reaches a maximum at the blocking temperature $T_B$. Above $T_B$ the ZFC and FC curves are identical because the nanodot moments are superparamagnetic. As can be seen from Figure S2 $T_B$ increases with nanodot size. The measured dependence of $T_B$ on $t_{ND}$ is shown in Figure 1d.

When an array of magnetic nanodots has a uniform size distribution there is a single blocking temperature and below $T_B$ the FC magnetization becomes constant (ref. 19). Clearly, this is not the case for the FC data shown in Figure S2 (a-c) for which the FC magnetization increases significantly below $T_B$. Such an increase in the FC magnetization indicates that there is a distribution of nanodot blocking temperatures and, thereby, nanodot sizes. In this case $T_B$ corresponds to the blocking temperature of the largest nanodots. As the temperature is decreased increasing numbers of nanodots consequently become blocked but because of reduced thermal fluctuations they contribute a net larger magnetic moment than they would have if they were blocked at a higher temperature. Thus the FC magnetization increases below $T_B$. As can be seen from Figure S2, when $t_{ND}$ ~1 nm, the FC magnetization data becomes nearly constant below $T_B$ indicating that nearly all the nanodot moments are blocked below $T_B$. As the nanodot layer thickness is reduced below ~1 nm the increase in the FC magnetization below $T_B$ becomes increasingly stronger.

**Influence of nanodot magnetic orientation on the TMR suppression in the Kondo regime**

The influence of the superparamagnetic behaviour of the nanodots on the suppression of the TMR in the Kondo regime was studied by considering the effect of cooling the MTJs in zero and finite magnetic field. As discussed in the previous section the magnetic ordering of the magnetic moments of the individual dots depends on the field in which the nanodots are cooled. For zero field cooling the net magnetic moment of the ensemble of nanodots is nearly zero (see Figure S2). Figure S3 compares the conductance versus voltage curves at 10 K of an MTJ of the form 10 MgO/ 15 Ta/ 4 CoFe/ 2.3 MgO/ 0.5 CoFe/ 2.3 MgO/ 15 Ta (thicknesses in nm) when the device is cooled in zero field (ZFC) and in a 1T field (FC). Magnetic fields of 1T and -0.05T are applied to set the state of the magnetic electrodes in either the parallel (P) or the antiparallel (AP) configurations, respectively. In both cases a peak in conductance is observed at zero bias voltage with similar half-widths at half maximum but the magnitude is slightly different as can be seen from the related TMR versus voltage curves shown in the figure. The TMR is suppressed near zero bias for the same range of voltages for which the conductance is enhanced independent of the cooling field. However, the magnitude of the TMR is strongly affected by the cooling field and, presumably, the nanodot ensemble magnetic configuration. The maximum TMR observed is 9.8% and 6.1 %, respectively for the FC and ZFC data but the ratio of the maximum TMR value to that at zero bias is nearly the same. In both cases the zero bias TMR is ~58% of the maximum TMR value.

**Replots of the data in Figure 4 normalized at bias=0.082 V**

The plots in Figure 4 (a,b, and c) are replotted in Figure S4 with the data normalized at the bias voltage of 0.082 where the color represents the ratio of TMR to $TMR_{bias=82mV}$ for each $t_{ND}$. The color represents different normalized TMR scales in each part.

**3-dimensional plots of the data in Figure 2 and 4**

The 2-dimensional contour plots in Figure 2 (a,b,d,e,g, and h) and 4 are replotted in Figure S5 and S6 as 3-dimensional contour plots where the color represents resistance in the parallel state or TMR. The 3-dimensional plot in Figure S6 shows the raw data without interpolation.

**References for supplementary information**

(S1) T. B. Tran *et al.*, *Phys. Rev. Lett.* **95**, 076806 (2005).

(S2) L. Yan *et al.*, *Appl. Phys. Lett.* **88**, 142901 (2006)

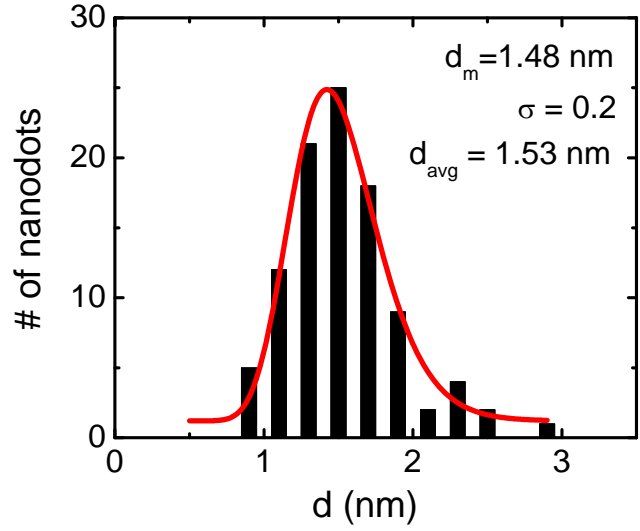

**Figure S1**. Histogram of nanodot diameters measured from the through foil TEM image shown in Figure 1b for the case of $t_{ND}$ = 0.25 nm. The solid red line is a fit of a log-normal distribution function $f(d) = N/(\sigma d\sqrt{2\pi})\exp[-\ln^2(d/d_m)/(2\sigma^2)]$, where $d$ is the nanodot diameter, $d_m$ ~1.48 nm, and $\sigma$ ~ 0.2 is the standard deviation of $f(d)$. The average diameter, given by $d_{avg} = \frac{1}{n}\sum_{i=1}^{n} d_i f(d_i)$ is ~1.53 nm.

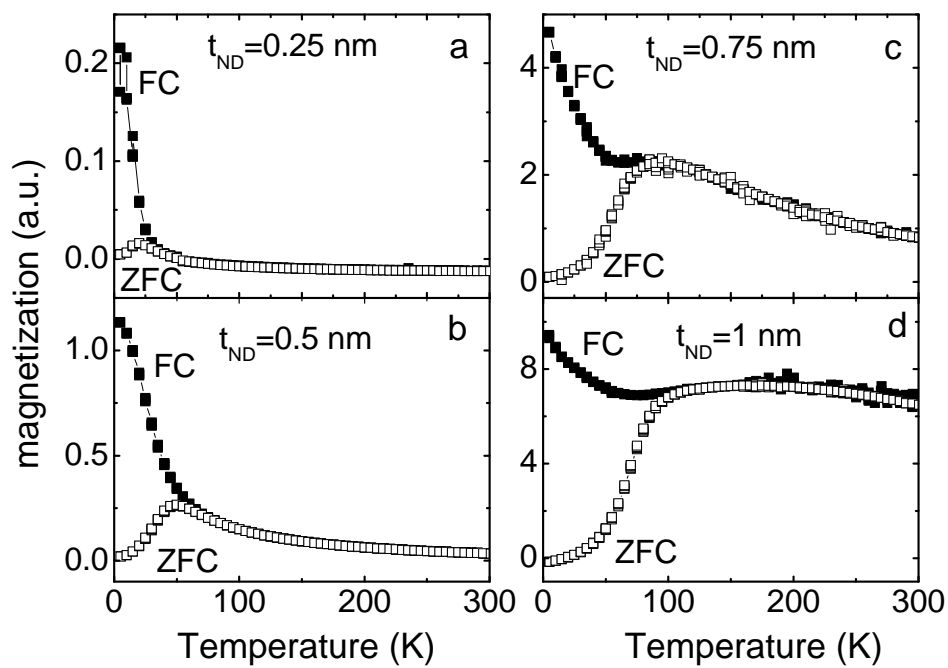

**Figure S2**. Field cooled (FC: solid symbols) and zero field cooled (ZFC: open symbols) magnetization versus temperature curves for $t_{ND}$= 0.25, 0.5, 0.75, and 1 nm.

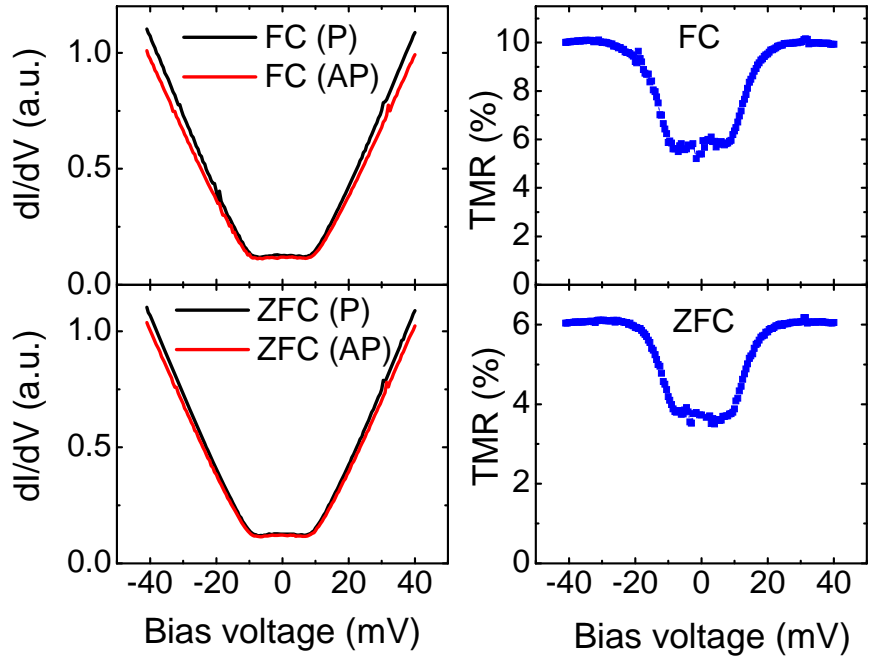

**Figure S3.** The conductance and TMR at 10K of an MTJ structure of the form 10 MgO/ 15 Ta/ 4 CoFe/ 2.3 MgO/ 0.5 CoFe/ 2.3 MgO/ 15 Ta (thicknesses in nm). The solid black and red lines correspond to the conductance for the parallel (P) and antiparallel (AP) configuration of the magnetic electrodes, respectively. The blue data points correspond to values of TMR derived from the P and AP conductance versus voltage curves for the FC and ZFC data.

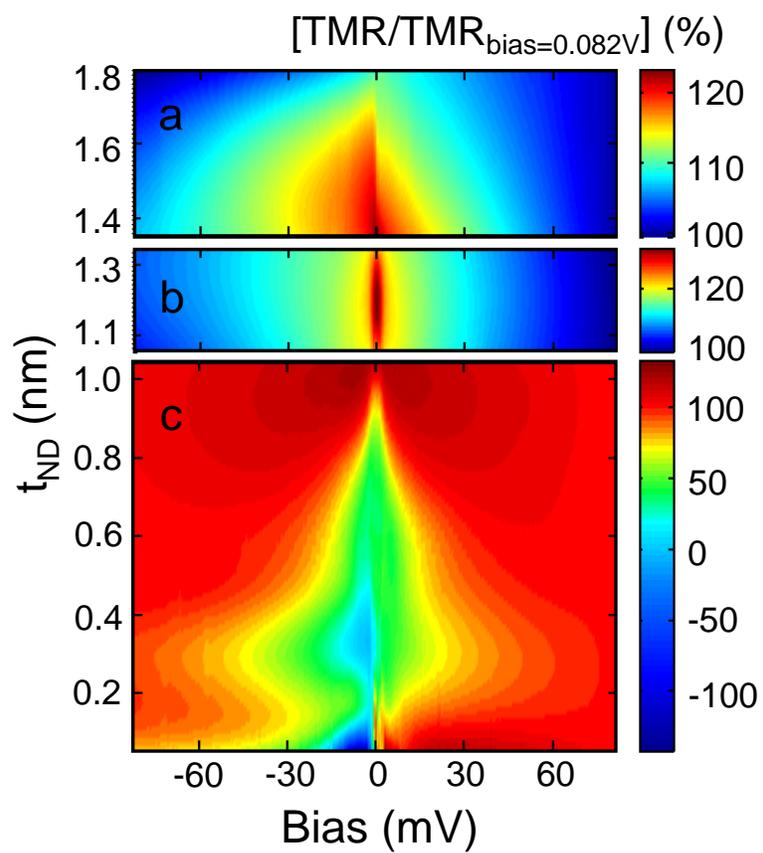

**Figure S4.** Normalized plot of Figure 4 at bias voltage 0.082V for each $t_{ND}$.

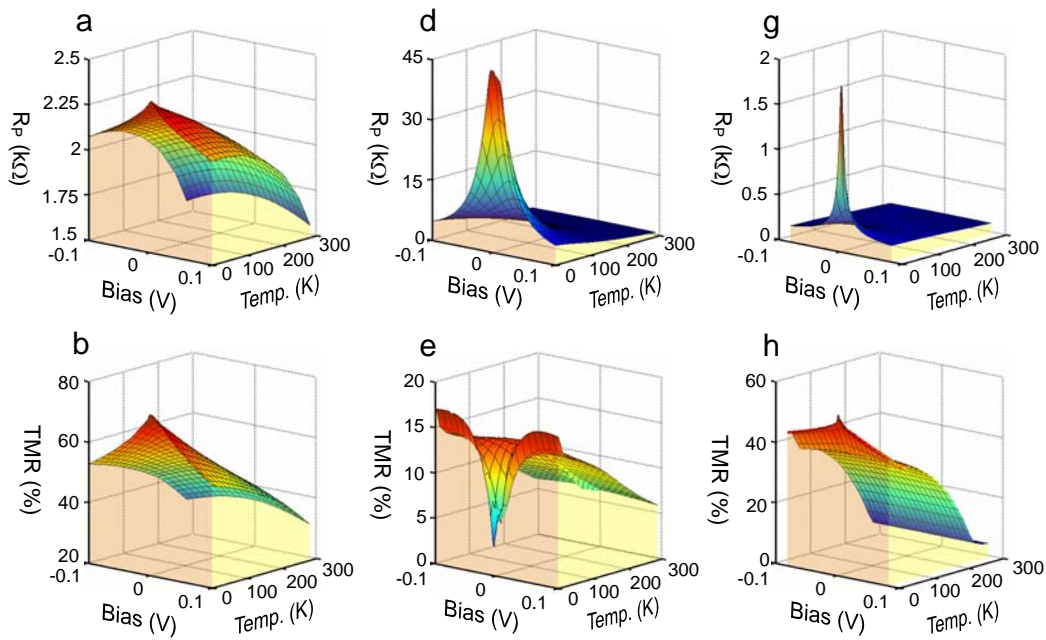

**Figure S5.** 3D plots of the data from Figure 2 (a,b,d,e,g, and h).

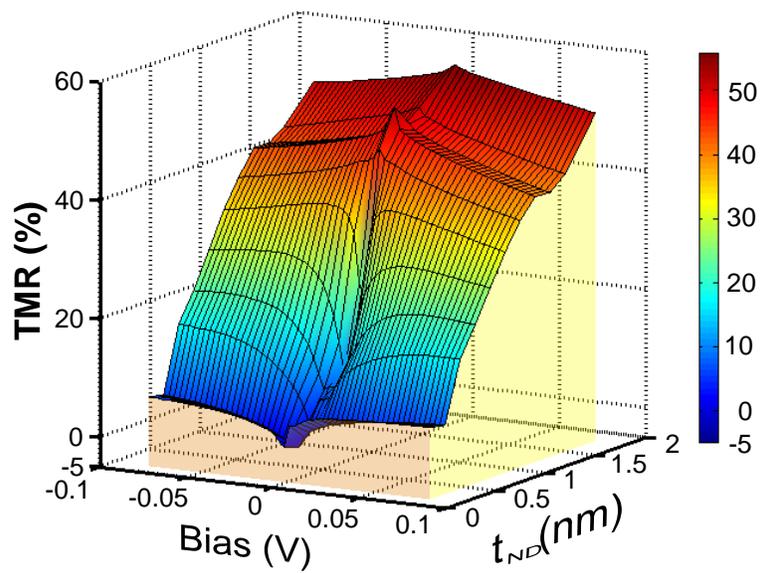

**Figure S6.** Data from Figure 4 are replotted as a 3D plot.